# Machine Learning and Deep Learning – A review for Ecologists

Maximilian Pichler[1,*], Florian Hartig[1]

[1] Theoretical Ecology, University of Regensburg, Universitätsstraße 31, 93053 Regensburg, Germany

* corresponding author, maximilian.pichler@biologie.uni-regensburg.de

*Keywords:*

Artificial intelligence, machine learning, deep learning, big data, causal inference

## Abstract

1. The popularity of Machine learning (ML), Deep learning (DL), and Artificial intelligence (AI) has risen sharply in recent years. Despite this spike in popularity, the inner workings of ML and DL algorithms are often perceived as opaque, and their relationship to classical data analysis tools remains debated.
2. Although it is often assumed that ML and DL excel primarily at making predictions, ML and DL can also be used for analytical tasks traditionally addressed with statistical models. Moreover, most recent discussions and reviews on ML focus mainly on DL, missing out on synthesizing the wealth of ML algorithms with different advantages and general principles.
3. Here, we provide a comprehensive overview of the field of ML and DL, starting by summarizing its historical developments, existing algorithm families, differences to traditional statistical tools, and universal ML principles. We then discuss why and when ML and DL models excel at prediction tasks and where they could offer alternatives to traditional statistical methods for inference, highlighting current and emerging applications for ecological problems. Finally, we summarize emerging trends such as scientific and causal ML, explainable AI, and responsible AI that may significantly impact ecological data analysis in the future.
4. We conclude that ML and DL are powerful new tools for predictive modeling and data analysis. The superior performance of ML and DL algorithms compared to statistical models can be explained by their higher flexibility and automatic data-dependent complexity optimization. However, their use for causal inference is still disputed as the focus of ML and DL methods on predictions creates challenges for the interpretation of these models. Nevertheless, we expect ML and DL to become an indispensable tool in E&E, comparable to other traditional statistical tools.

## Introduction

In recent years, Machine learning (ML), Artificial intelligence (AI) and Deep learning (DL) have revolutionized almost all areas of science (Jordan & Mitchell, 2015). Early ML algorithms emerged together with the first computers in the '50s, and co-evolved with the advancement of computing power ever since. During the '90s, the ML field experienced its first bloom, when a wave of fundamental concepts and algorithms such as boosting, bagging, shrinkage estimation and random forest were discovered. These algorithms challenged, for the first time, the supremacy of classical probability-based statistical models for data analysis and predictions. In the last decade, a second revolution occurred with the rediscovery of deep neural networks, fueled by the availability of graphical processing units (GPUs; 'graphic cards') who made applying these large neural networks practical for the first time. Famous break-throughs of DL include playing Go (AlphaZero, Silver et al., 2017), natural language processing (NLP, e.g. GPT-2, Radford et al., 2019), detecting and identifying objects in images (Mask R-CNN, He et al., 2017), and predicting protein structures (AlphaFold, Jumper et al., 2021).

Research in ecology and evolution (E&E) has eagerly adopted both waves of innovation. Several recent reviews highlighted the potential of the more novel DL approaches (Borowiec et al., 2022; Christin et al., 2019; Tuia et al., 2022; Wäldchen & Mäder, 2018), in particular for processing ecological data such as species recognition from video and audio analysis (Mac Aodha et al., 2018; e.g. Fritzler et al., 2017; Gray et al., 2019; Guirado et al., 2018; Lasseck, 2018; Tabak et al., 2019) or for extracting trait or behavioral information (Dunker et al., 2020; Graving et al., 2019; Mathis et al., 2018; Ott & Lautenschlager, 2021; Pereira et al., 2019). A second area where both traditional ML and DL approaches are already widely used in E&E is predictive modelling. Examples include filling missing links in networks (e.g. Desjardins-Proulx et al., 2017), as part of or together with traditional mechanistic models (Rammer & Seidl, 2019; Reichstein et al., 2019), by replacing parts of or approximating differential equations with neural networks (R. T. Q. Chen et al., 2019; Rackauckas, Ma, et al., 2021), or for species distribution models (D. Chen et al., 2018; Elith & Leathwick, 2009; Harris et al., 2018; Wilkinson et al., 2019).

Despite their rising popularity, however, the principles and inner workings of ML and DL algorithms are still perceived as opaque by many researchers in E&E, and their relationship to more classical tools of data analysis, in particular statistical models, remains debated. Trained ML and DL models are often called a “black box” because, due to their complexity, it is difficult to know what they have learned or how they make predictions. To address this problem, the field of explainable AI (xAI) develops methods for understanding the properties of trained ML or DL models (Ribeiro et al., 2016; Ryo et al., 2021). Moreover, a pervasive concern is that ML models are trained for prediction, but the best predictive model must not necessarily correspond to the causal model (Breiman, 2001b; Pearl, 2021; Pearl, 2019; Box 4). Many researchers thus assume that ML and DL are unable to generate ecological understanding and thus can only be used as predictive tools. This view, however, neglects that there is active research to expand ML and DL methods also to causal inference (Schölkopf, 2019), which is the classical domain of inferential (causal inference, confirmatory and similar) statistics.

A second reason for confusion about the field is the wealth of algorithms that have been developed in recent years. Most recent reviews on ML have exclusively focused on DL (Borowiec et al., 2022; Christin et al., 2019; Wäldchen & Mäder, 2018). As these algorithms differ considerably in their behavior to simpler, more traditional algorithms such as k-nearest-neighbor or boosted regression trees, not all statements that are made with respect to DL algorithms apply across the field of ML algorithms in general. For example, image based tasks such as automatic species identification (e.g. Ferreira et al., 2020; Tabak et al., 2019) profit from the use DL algorithms because they can better process spatial patterns than other ML algorithms (LeCun et al., 2015), whereas traditional ML algorithms often cope better with lower number of observations (e.g. Pichler et al., 2020) or structured (tabular) data (cf. Arik & Pfister, 2020).

Third, a narrow focus on specific algorithms often prevents researchers to appreciate how general principles apply or change across all ML and DL algorithms. For example, the general principles of regularization via shrinkage and model averaging form the backbone of nearly all ML and DL models. Other principles must be re-learned when going from the classical ML to DL. For example, the bias-variance trade-off classically predicts that rising the model complexity reduces systematic model error (bias) at the cost of increasing uncertainty (variance) of the parameters and thus the risk of flawed predictions for new observations (Box 3). For DL models, however, it was shown that for achieving low generalization error (accurate predictions for new observations), networks have to be overparametrized (Frankle & Carbin, 2019; Huh et al., 2021; S. Zhang et al., 2021). The question why deep neural networks do not suffer from overparameterization but even depend on it is still heavily debated (Sejnowski, 2020), and we will comment later more on it.

Here, our aim is to provide a comprehensive overview of the principles of ML and DL in their full breath for ecologists, starting with their historical developments, how these tools differ from traditional statistical tools and how they can be applied for predictive and explanatory modeling (for recent reviews on specific methods, see e.g. DL: Borowiec et al., 2022; Christin et al., 2019; Wäldchen & Mäder, 2018; Computer vision: Lürig et al., 2021 or for specific application areas of ML, see e.g. Tuia et al., 2022 (wildlife management)). After discussing the algorithmic ideas, history and general properties of ML and DL algorithms, we focus on understanding the general mechanisms that make ML and DL excel in certain predictive and analytical tasks and how this can be used by ecologists. We also discuss the current limitations of ML and DL and where traditional statistical methods are preferable, highlighting also current and emerging applications for ecological problems.

# History of ML and DL and its relation to statistics

## Statistics as the starting point for ML

The roots of ML and DL go back a long way, and the development of this field is tightly connected to the development of statistics. It is therefore useful to place the field in the context of the development of modern statistics.

Apart from Bayesian statistics, many foundational statistical principles such as the maximum likelihood estimate (MLE) or null hypothesis significance testing (NHST) were established in the first half of the 20$^{th}$ century (Fig. 1, left). In the core of these classical parametric methods is the idealization of a data-generating model, which allows calculating the probability of the observed data, given the model assumptions and parameters (likelihood). Based on this model, eminent statisticians such as Fisher, Neyman and Pearson developed the theory and practice to estimate model parameters and calculate confidence intervals (CI) and p-values that has dominated data analysis in E&E until today (but see Dushoff et al., 2019; Gelman & Loken, 2014; Hartig & Barraquand, 2022; Muff et al., 2022).

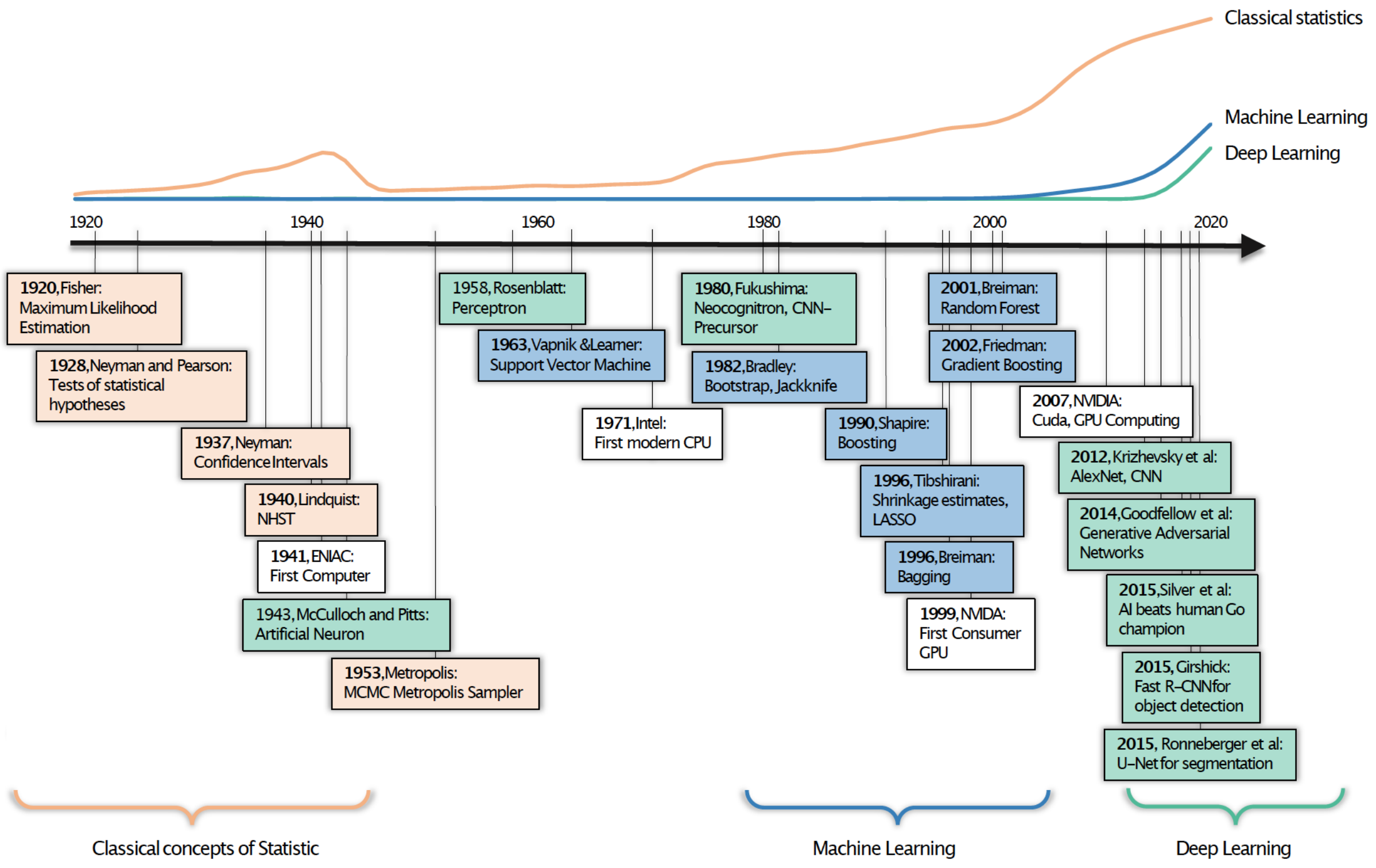

**Figure 1:** The three eras of statistical learning. The classic concept of statistic such as the maximum likelihood estimator (MLE), null hypothesis significance testing (NHST), or the Markov-chain-Monte-Carlo (MCMC) metropolis sampler were developed in the 1920-1940s. Common machine learning algorithms or techniques such as boosting, random forest, or the LASSO were discovered between 1980 and the early 2000s. While the theoretical foundation for Deep learning was postulated in the '60s, it has only gained popularity in recent years. The trend lines above the timeline correspond to frequency of the occurrence of classical statistics (orange), Machine Learning (blue), and Deep Learning (green) terms in scientific literature (see Appendix S1.1 for more details).

Initially, the data-generating model underlying these methods had to be relatively simple to make the calculations of the involved probabilities tractable. The emergence of the first computers (Fig. 1) made it possible to expand the complexity of parametric statistical models, also helped by the discovery of new numerical algorithms (e.g. Markov-chain-Monte-Carlo,

MCMC, 1958). In recent years, ecologists have steadily moved to more complex statistical models (Clark, 2005). Even so, when considering the known complexity of the natural world (Grimm et al., 2005), statistical models tend to be rather simple and rigid, due to the mathematical difficulties involved in calculating likelihoods for more complex or flexible models, and it is an important caveat of tradition statistical methods that the quality of their inference is conditional on the model assumptions.

## Box 1: Basics of Machine Learning

### General objective of ML

The objective of ML is to build a good predictive model. By "good", we mean that the model should predict well to new data. Sometimes ML models make almost no errors on the data they are trained on but fail for new data (we say the model overfits). A more complex and flexible model has a higher risk of overfitting. The trade-off between complexity and flexibility can be depicted by the bias-variance tradeoff (Box 3). The general ideal of ML algorithms is thus to take a certain algorithmic structure and then adjust their parameters to the data (training), while simultaneously adapt its complexity by optimizing the bias-variance trade-off so that the fitted model generalizes well to new data.

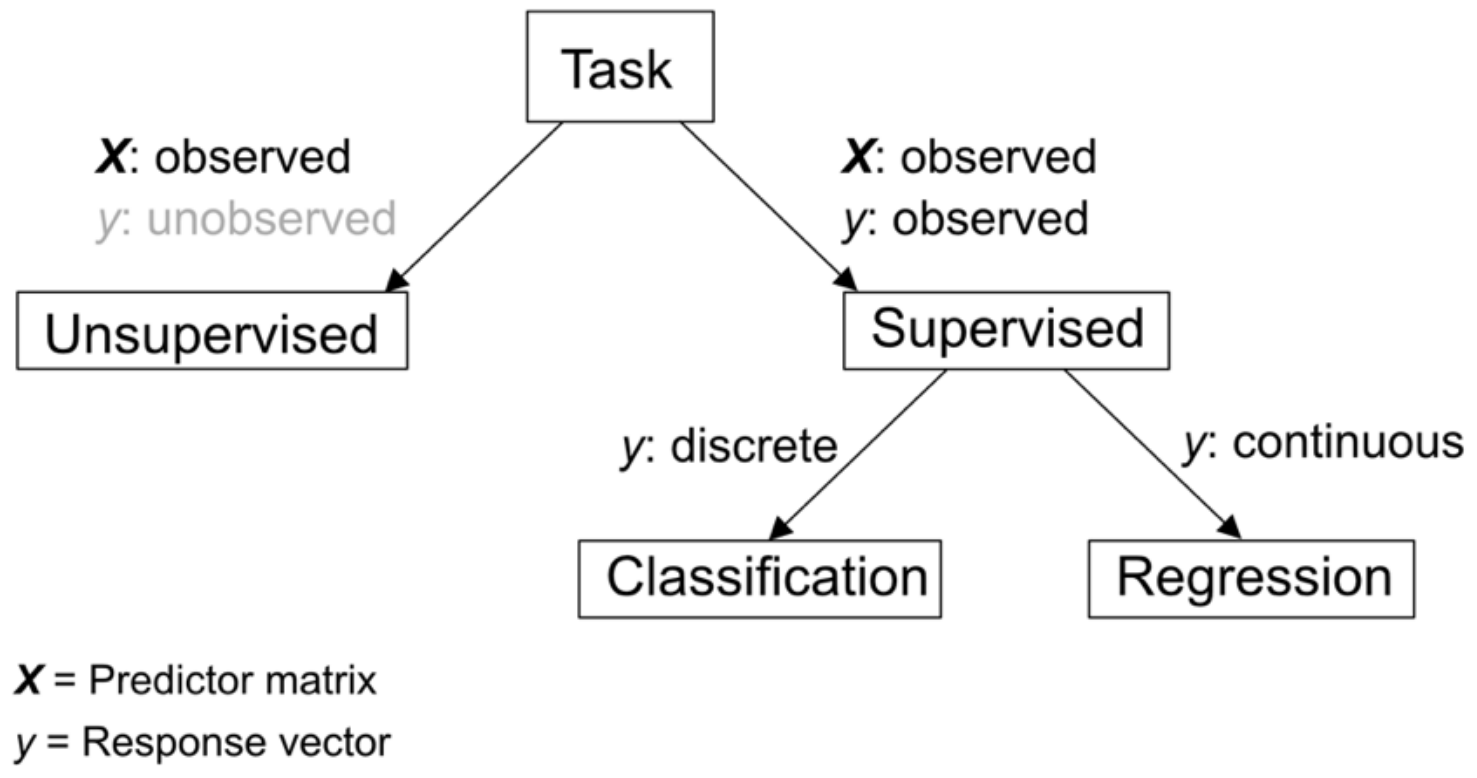

**Figure I:** Decision tree to assist in task identification. Given feature matrix ***X*** and a response vector *y*, the first decision is to choose between unsupervised (outcome y is unobserved) and supervised (outcome y is observed) learning. In the case of supervised learning, if *y* is discrete (e.g., species classes), it is a classification task, and if *y* is a continuous variable (e.g. biomass), it is a regression task.

### Tasks and learning situations

In ML, the different use cases for the algorithms are called tasks. In **supervised learning**, the algorithm is presented with examples for the "correct" execution of the task, and the model is trained to minize the differences its own and the "correct" actions. Common supervised tasks are **classification** (e.g. labeling of images) and **regression** (prediction of a numeric variable). In constrast to that, **unsupervised learning** refers to tasks where no examples are supplied, and algorithms optimize some general loss function (e.g. genomic species delimitation, see Derkarabetian et al., 2019). Finally, in **reinforcement learning**, the ML algorithm is trained by interacting with a (virtual) environment. Reinforcement learning is used in tasks in which the learning depends on executed actions and their produced consequences, for instance, playing strategy computer games such as DOTA (OpenAI et al., 2019) or Starcraft (Vinyals et al., 2019).

**Model classes and architectures**

In principle, any algorithm that makes predictions for a certain task can be used for ML. In practice, for **supervised learning**, model classes and architectures that are commonly used can broadly be divided into neural networks, which mimic the functioning of a brain, regression and classification trees, and distance-based method (see section "Important ML and DL algorithms in more detail"). **In unsupervised learning**, model classes can be broadly divided into agglomerative hierarchical methods and methods where the number of clusters have to be specified a priori (e.g. k-means) (Box 2).

**Training the models**

In **supervised** and **reinforcement learning**, training a model consists of two steps. The first step is the definition of a loss function, which measures the current score (performance) of the algorithm in solving a certain task. The loss function differs for classification and regression tasks (e.g., mean squared error and categorical cross-entropy are common loss functions for regression and classification tasks). The second component is the optimizer, which updates parameters of the algorithm with the goal to increase its performance. In **unsupervised learning** it depends on the method, but a common approach is to use similarities between observations to decide if observations are grouped together or not.

## Machine learning

The rising availability of computers around the 1980s allowed not only more refined numerical solutions for classical statistical methods, but also the development of alternative modelling approaches for data analysis and predictions that we collectively refer to as "machine learning". Although these approaches differ in detail, we see their communality in the realization that abandoning the data-generating model (connected to the ability to calculate p-values, CIs and all that) in favor of generic algorithmic structures that can be trained to data often achieves lower errors for predictive tasks (for general ML principles, see Box 1) (Breiman, 2001b; Shmueli, 2010). Examples of early ML algorithms are neural networks (McCulloch & Pitts, 1943), random forest (Breiman, 2001a), and boosted regression trees (Friedman, 2001) (more on these in the section 'Important ML Algorithms in more detail').

The algorithmic nature of the new ML models lacked the necessary distributional assumptions for calculating p-values and CIs and fueled the development of non-parametric approaches for estimating model uncertainty. A famous example is the bootstrap (Efron, 1992), a resampling technique that is often used for estimating CIs on the parameters and predictions of statistical or ML models. Another example is cross-validation, where a part of the data is used to train the model and the other part of the data is used to assess the error (Stone, 1974; see Roberts et al., 2017 for cross-validation strategies for structured ecological data). Since either of these methods require repeated evaluations of the model, their application would be unimageable without computers and can even today be computationally challenging for complex models (more on this in section 'Why does machine learning work').

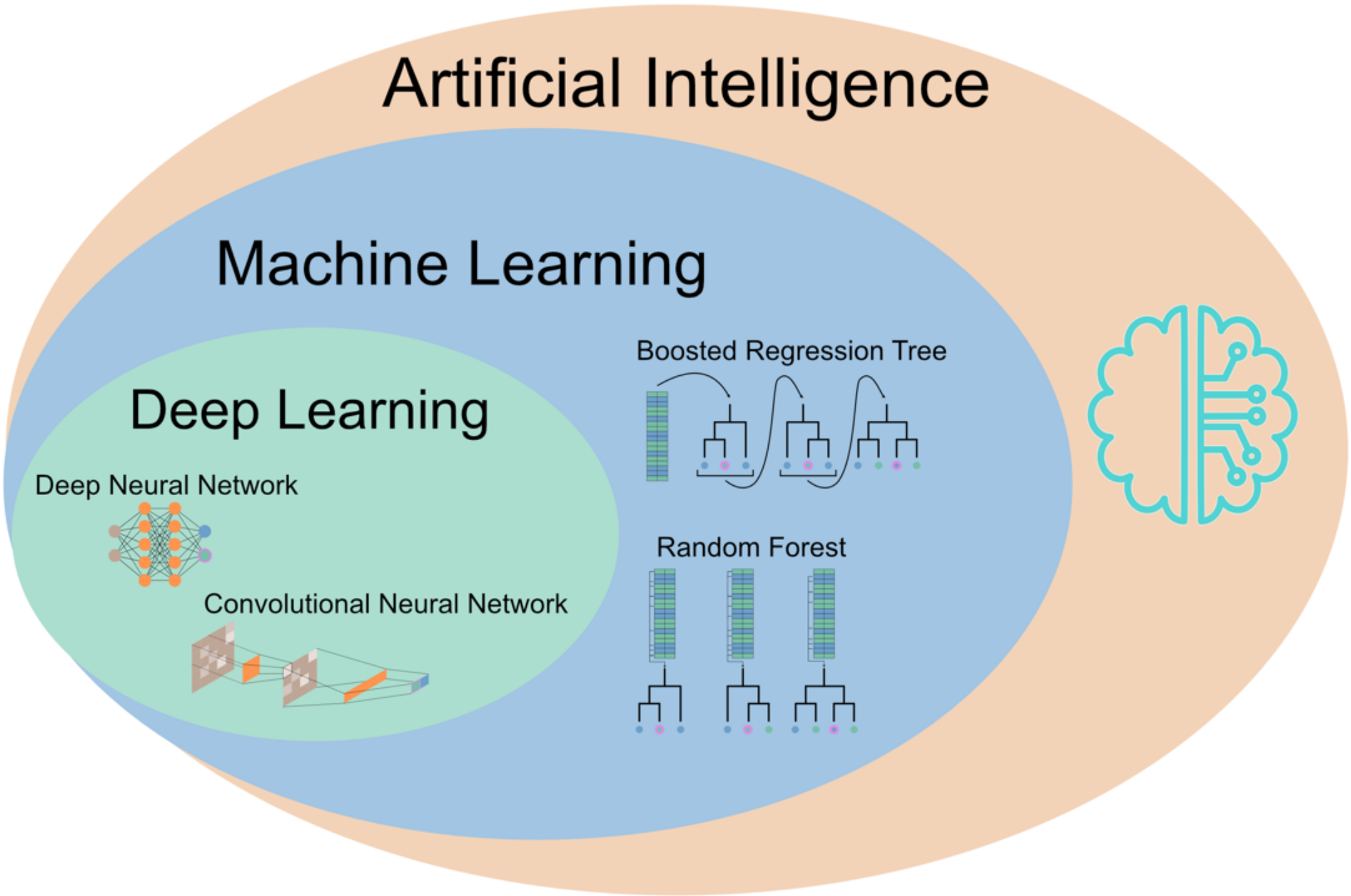

**Figure 2:** Relationship between Artificial Intelligence (AI), Machine Learning (ML), and Deep Learning (DL). AI refers to algorithms that are capable of achieving similar to human performance in specific decision or recognition tasks. This is sometimes contrasted to the aspiration to achieve Artificial general intelligence (AGI), referring to AI algorithms that can perform a wide range of tasks and may display human-like abilities in cognitive tasks such as reasoning, logic or common sense. ML algorithms serve as a tool for AI systems to learn from data and make a decision based on data. There are many different ML algorithms such as Boosted Regression Trees (BRT) or Random Forest (RF). Within ML, a family of ML algorithms based on artificial neural networks emerged

in recent years. Due to their similarities in terms of functioning and their backbone, DL is considered as a family of its own.

## Deep learning

The co-evolution of computational resources and ML algorithms experienced a final peak with the emergence of DL algorithms in the last decade. DL algorithms are neural networks (McCulloch & Pitts, 1943) that differ from classical artificial neural networks mainly by their size. While many algorithms and network architectures that are used today were already described in the '80s and '90s (e.g. Fukushima, 1980; Lecun et al., 1998), their practical application was prevented by the lack of computing power at the time. This changed with the emergence of graphical processing units (GPU) in the '90s (Fig. 1). Although GPUs were originally developed for computer games or other graphical rendering tasks, it was quickly realized that they are often far more efficient than CPUs for certain numeric and linear algebra tasks. Krizhevsky et al., 2012 ushered in the new era of DL when they demonstrated that their competition-winning neural network could be trained within hours on a GPU instead of days or weeks using a CPU. Today, large DL models trained on GPUs with hundreds of millions parameters dominate the competition for many complex ML tasks, and their behavior often differs markedly from simple ML algorithms (see section "Why ML works").

## Box 2: Unsupervised learning in E&E

Also unsupervised learning algorithms (for definitions, see Box 1) have interesting applications in E&E. In most cases, the goal is to find patterns in the feature space, for example to reduce the dimensionality of the data, finding clusters of similar data, and to detect anomalies (Fig. II).

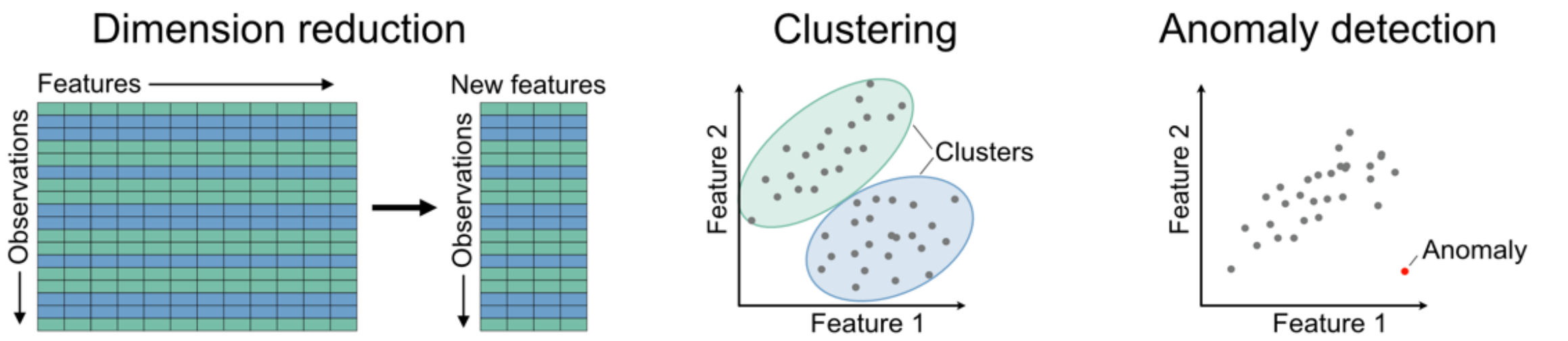

- Principal component analysis (PCA)
- t-distributed stochastic neighbor embedding (tSNE)
- (Variational) Autoencoder

- k-means
- density based clustering (e.g. DBSCAN)
- hierarchical clustering
- Unsupervised random forest

- (Variational) Autoencoder
- Clustering algorithms (e.g. k-means)
- k-nearest-neighbor
- Support vector machines

**Figure II:** The three main tasks and their algorithms in unsupervised learning. Dimension reduction techniques reduce the dimension of the data by discarding redundant or non-task relevant information. Clustering algorithms can identify patterns in the data which correspond to mechanistic processes. Anomaly detection is used to identify observations that may have originated form a different data generating process.

Examples for algorithms that perform dimension reductions that are well-known to ecologists include ordination methods such as principal component analysis (PCA) or t-distributed stochastic neighbor embeddings (tSNE), but also deep learning algorithms such as variational autoencoders can be used for dimension reductions, and the latter also works for more complex data such as pictures. The same is true for clustering and anomaly detection tasks: on top of simple methods such as k-means, which should be familiar to many ecologists, there are now deep-learning methods available that generally have advantages if the data is highly structured, such as in images.

# Important ML and DL algorithms in more detail

Considering that ML branched off from classical statistical models with the goal of increasing model flexibility and complexity while at the same time abandoning the idea of a probabilistic model, it seems obvious to discuss advantages and disadvantages of this decision. We will do so in section "Why ML works".

Before that, however, it will be useful to explain the most important ML algorithm in somewhat more detail. In the main text, we concentrate on algorithms for supervised learning (see Box 1 for definitions of ML tasks), but we also provide a short overview about unsupervised learning in Box 2. Note that classical statistical models such as linear and logistic regression models can also be used for supervised regression and binary classification tasks, respectively. Arguably, they provide a baseline that machine learning models should be able to beat. Because we assume that ecologists are aware of these models, however, and because our very aim here is to understand why ML algorithms can beat these models, we do not describe them in this section. R, Python, and Julia code examples for all ML and DL algorithms that are discussed (Table 1) are available in the Supporting Information S1.2 or at https://maximilianpi.github.io/Pichler-and-Hartig-2022.

**Table 1:** Overview of common supervised machine learning algorithms with their most common application areas. Word clouds were created by searching abstracts and titles in the ecology and evolution literature within the specific ML algorithms for ecological keywords, size of words corresponds to their frequency (see Appendix S1.1).

| Machine learning algorithms | Description | Data Type | Application areas |
|---|---|---|---|
| Lasso, Ridge Regression:<br>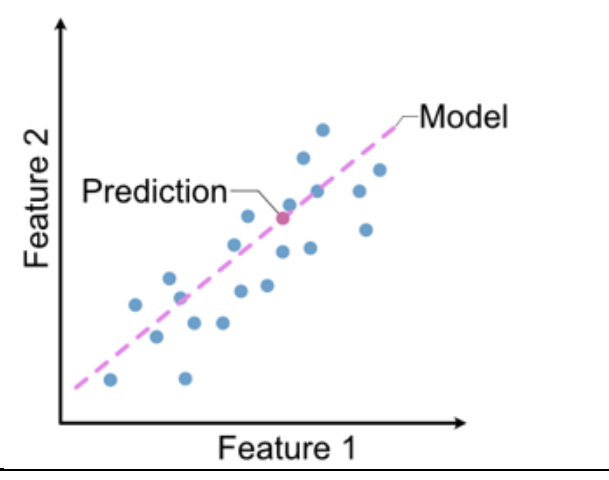<br> | Regression models with regularized coefficients (Appendix S1.2.1):<br>+ highly interpretable<br>+ few observations<br>- limited flexibility | Tabular data:<br>- Classification<br>- Regression | 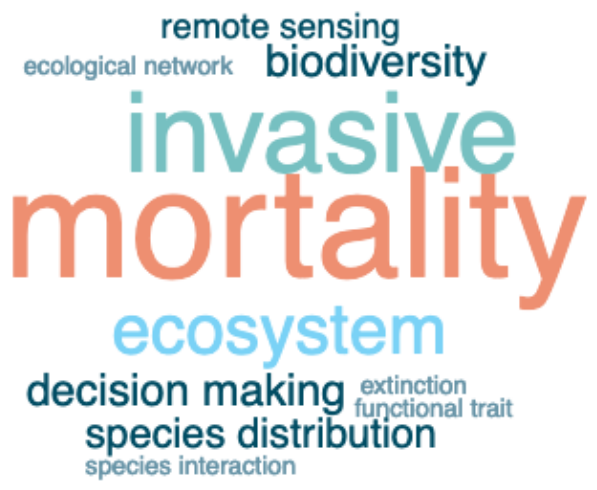<br> |
| Support vector machines:<br>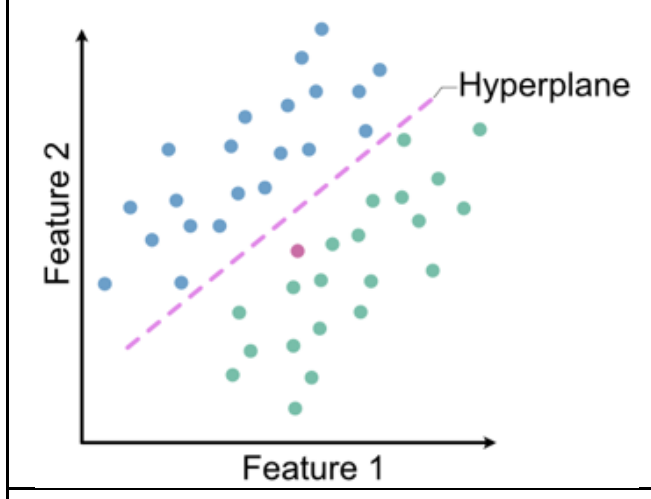<br> | Hyperplane is optimized to separate response classes (Appendix S1.2.2):<br>+ fast and memory efficient<br>+ high dimensional data<br>- kernel dependent<br>- no probabilities | Tabular data:<br>- Classification<br>- Regression | 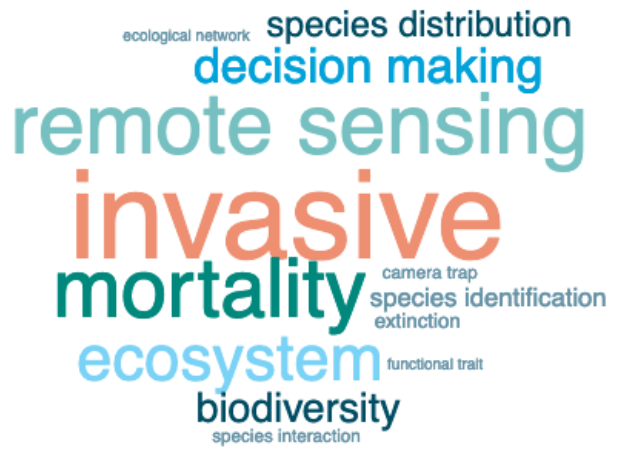<br> |
| k-nearest-neighbor:<br>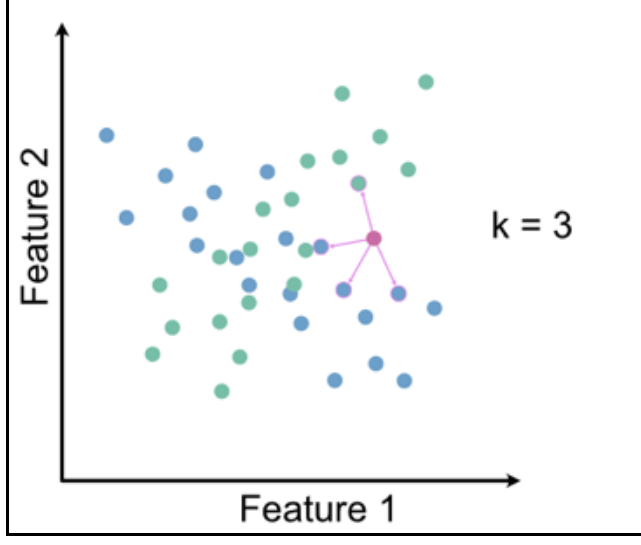<br> | K nearest neighbors in feature space decide response (e.g., by majority voting) (Appendix S1.2.3):<br>+ simple<br>+ no training<br>- scales poorly<br>- high dimensionality | Tabular Data:<br>- Classification<br>- Regression | 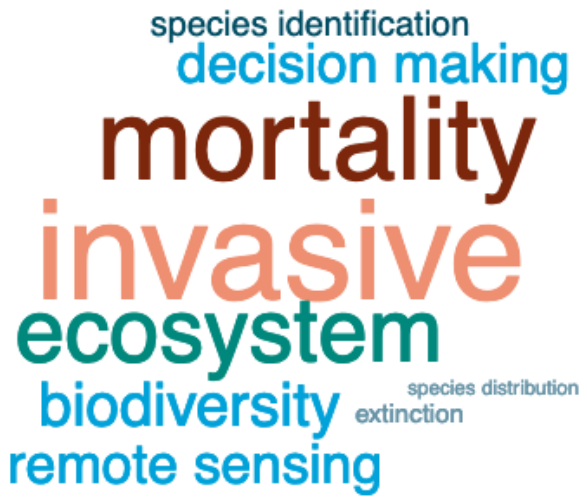<br> |
| Random Forest:<br>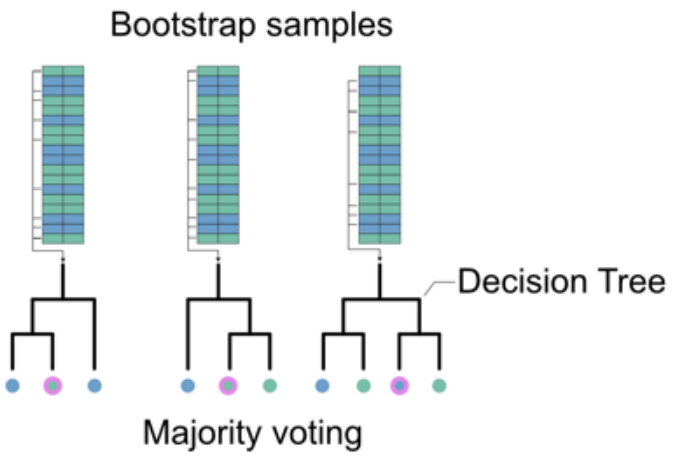 | N decision (regression) trees are fitted on bootstrap samples. Split variable is selected from random subset of variables (Appendix S1.2.4):<br>+ flexible<br>+ robust (e.g. outliers)<br>+ few hyper-parameters<br>(+) variable importance<br>- scales poorly | Tabular Data:<br>- Classification<br>- Regression | 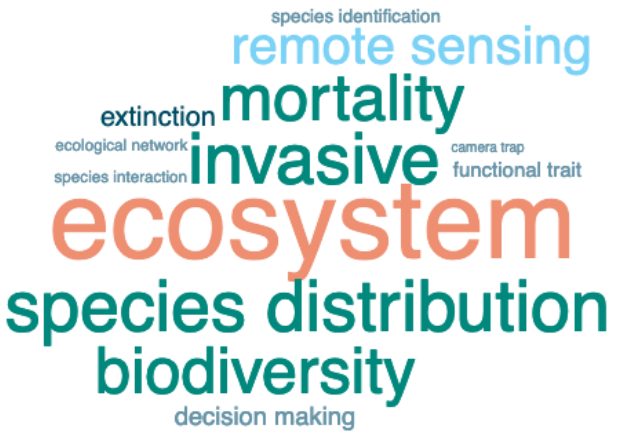 |
| Boosted Regression Trees:<br>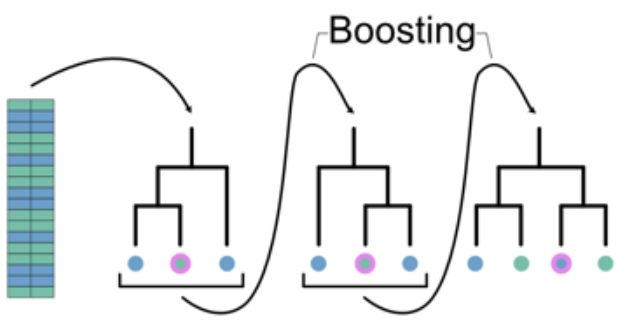 | N trees are fitted sequentially to minimize an overall loss function (Appendix S1.2.5):<br>+ flexible<br>(+) variable importance<br>- many hyper-parameters<br>- high complexity | Tabular Data:<br>- Classification<br>- Regression | 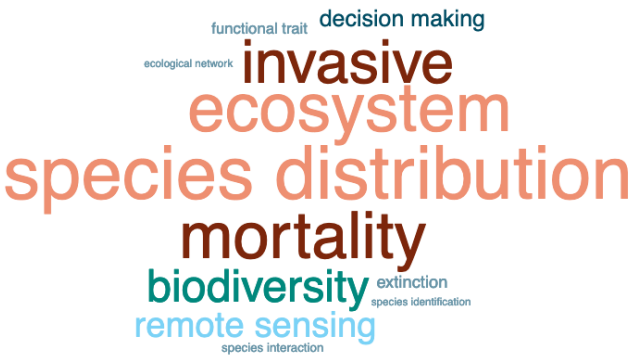 |
| Deep Neural Networks:<br>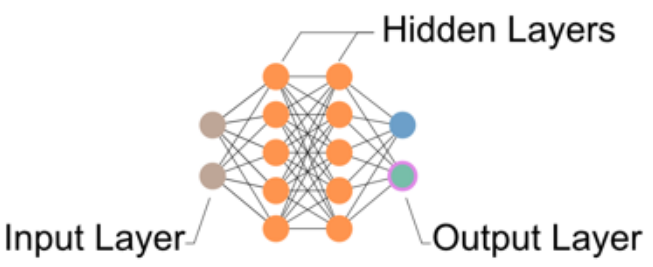 | Input (features) are passed through many hidden layers. Last layer maps into response space (Appendix S1.2.6):<br>+ flexible<br>+ adaptive to different tasks<br>- many hyper-parameters<br>- computationally expensive | Tabular Data:<br>- Classification<br>- Regression | 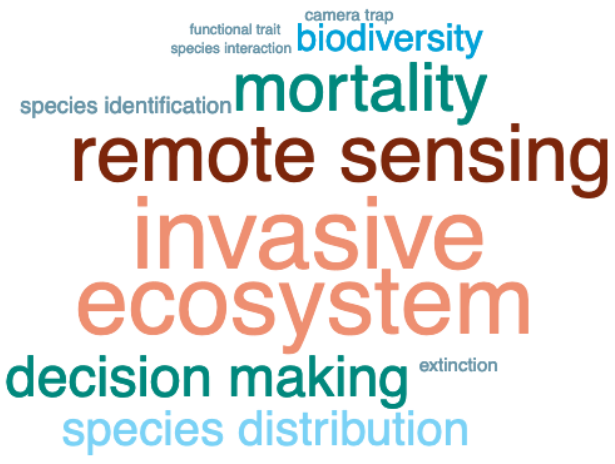 |
| Convolutional Neural Networks:<br>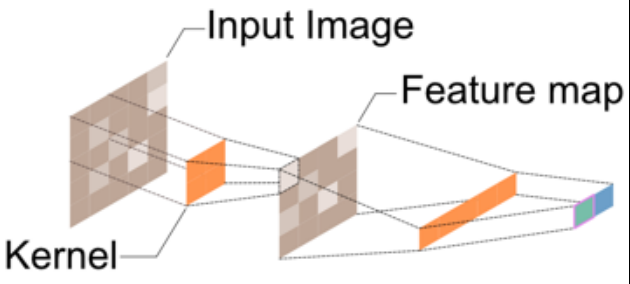 | Small kernels (filters) processes images before passing it to fully connected layers (Appendix S1.2.7):<br>+ flexible<br>+ detecting shapes and edges<br>- many hyper-parameters<br>- computationally expensive | Images:<br>- Classification<br>- Object detection | 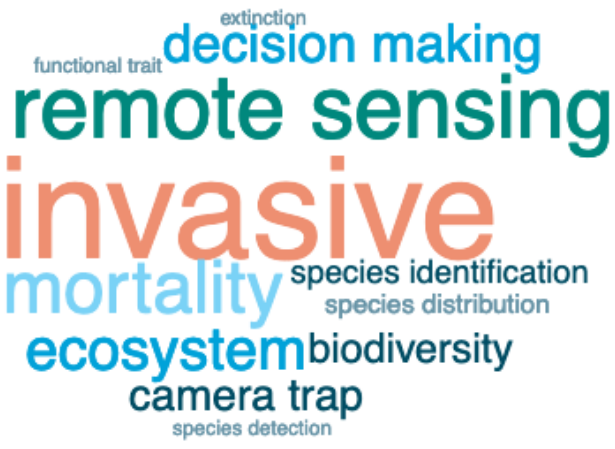 |
| Recurrent Neural Networks:<br>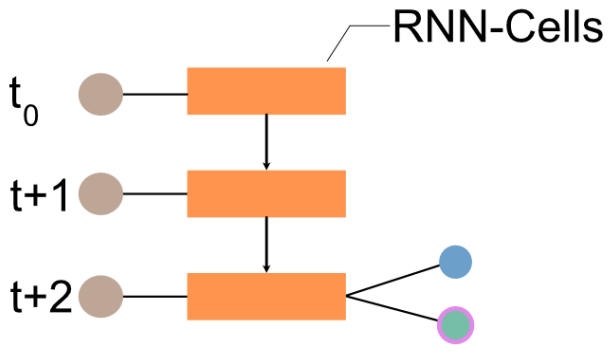 | RNN-Cells (e.g. Long short term memory cells) process the input sequence and hidden states are re-cycled (Appendix S1.2.8):<br>+ flexible<br>- long-term dependencies are difficult to learn<br>- many hyper-parameters<br>- computationally expensive | Sequences (e.g. temporal):<br>- Classification<br>- Regression | 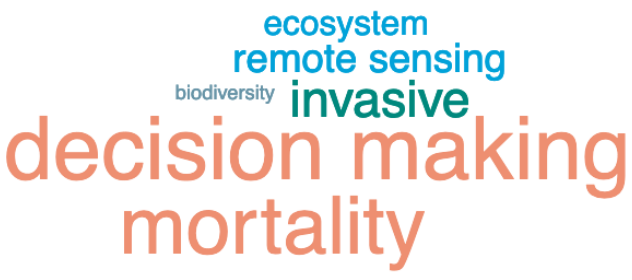 |
| Graph Neural Networks:<br>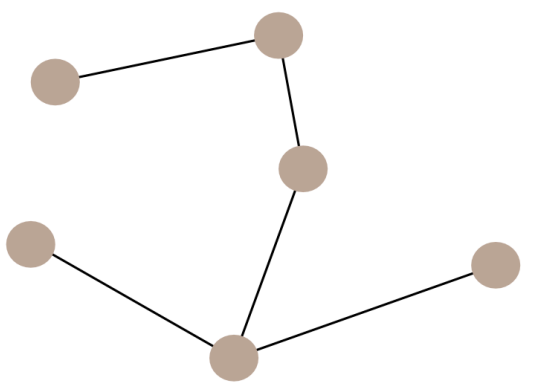 | GNNs operate directly on the edges and nodes of graphs. They can be used for a variety of different tasks such as node or edge classifications (Appendix S1.2.9):<br>+ flexible<br>+ non-Euclidean data<br>- many hyper-parameters<br>- computationally expensive | Graphs:<br>- Classification<br>- Regression | 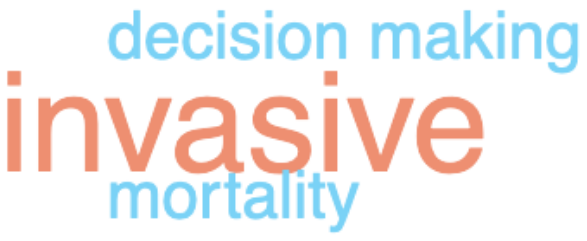 |
| | - complex data type | | |

**Table 2:** Common ML and DL libraries and frameworks.

| Name | Description | Language | Link |
|---|---|---|---|
| ranger | Random Forest algorithm. | R | https://github.com/imbs-hl/ranger |
| xgbooost | Boosted Machine framework. | R, Python | https://github.com/dmlc/xgboost |
| lightGBM | Boosted Machine framework. | R, Python | https://github.com/microsoft/LightGBM |
| caret | ML framework for hyper-parameter tuning and cross-validation. Supports different ML algorithms. | R | https://topepo.github.io/caret/ |
| mlr3 | ML framework for hyper-parameter tuning and cross-validation. Supports different ML algorithms. | R | https://mlr3.mlr-org.com/ |
| tidymodels | (ML) framework for hyper-parameter tuning and cross-validation. Supports different (ML) algorithms. | R | https://www.tidymodels.org/ |
| Scikit-learn | ML framework for hyper-parameter tuning and cross-validation. Supports different ML algorithms. | Python | https://scikit-learn.org/ |
| TensorFlow | Deep Learning framework. | R, Python | https://www.tensorflow.org/ |
| Keras | Higher-level deep learning framework. | R, Python | https://keras.io/ |
| PyTorch | Deep Learning framework. | R, Python | https://pytorch.org/ |
| PyTorch Geometric | Graph Neural Network (GNN) framework. Supports different GNN algorithms. | Python | https://github.com/pyg-team/pytorch_geometric |
| Flux | Deep learning framework. | Julia | https://fluxml.ai/Flux.jl/stable/ |
| MLJ | ML framework. Supports different ML algorithms. | Julia | https://alan-turing-institute.github.io/MLJ.jl/stable/ |

## Support vector machines

Already in 1963, Vapnik & Leaner proposed the generalized portrait algorithm, a predecessor of the support-vector machine (SVM). A SVM is a binary classifier that separates the available classes by a hyper-plane in the feature (predictor) space (see SVM in Table 1). The generalized portrait algorithm was computationally cheap to calculate (which was important at that time), but as the initial predecessor of artificial neural networks, the perceptron, it was unable to solve non-linear tasks. Later, the support vector machine overcame this obstacle by using a non-linear transformation of the feature space (the kernel-trick) to make the task linearly separable (Boser et al., 1992).

Prior the success of DL, SVMs were the most common method to classify images in E&E, particularly in remote sensing (e.g. Gualtieri & Cromp, 1999; Melgani & Bruzzone, 2004; Mountrakis et al., 2011). Their popularity was primarily fueled by their ability to efficiently deal with high dimensional data and by their relatively low requirements for the size of the training data (compared to DL).

## Ensemble models

With the ongoing advances in computing power in the 90's, a new ML paradigm emerged: Schapire (1990) showed that ensembles of weak learners (typically simple models such as linear regression models or classification and regression trees, see Friedman, 2001) can have a low predictive error if their predictions are averaged, even if each individual model has large predictive errors. This princinple of generating ensembles of "weak learners" gave rise to two prominent ML techniques: boosting and bagging.

Boosting is an enemble modelling approach where weak models are trained in sequence, either by training the next model to correct the errors of the previous model (AdaBoost , see Freund & Schapire, 1997), or by optimizing sequentially a general differentiable objective (cost) function, gradient boosting (Friedman, 2001) with the latter being state-of-the-art today (e.g. boosted regression trees (BRT) for species distribution models, see Elith et al., 2008; Elith & Leathwick, 2009).

In bagging (bootstrap aggregation) an ensemble of independent weak models is created by training models on bootstrap samples (Breiman, 1996). A famous representative is the Random Forest (RF) algorithm which additionally subsamples the features in each node of the decision trees (Table, 1; Breiman, 2001a).

Ensemble models are based on an important ML principle that is still valid today: simple algorithms or statistical models can be turned into more complex algorithms by creating ensembles, which are more difficult to interpret, but often have low prediction errors (see section "Why ML works"). BRT and RF are still widely applied, mostly for structured tabular data (Table 1), also because they cope better with smaller datasets than comparable DL models. Examples for recent applications of ensemble models in E&E include predictions in ecological networks (Pichler et al., 2020), linking gene variation to phenotypes (Brieuc et al., 2018), species distribution models (Elith & Leathwick, 2009), and various applications in remote sensing (Belgiu & Drăguţ, 2016).

## Neural networks

Possibly the most iconic ML architecture is a neural network, which is inspired by the architecture of our brains. The first fully functioning artificial neural network (ANN) was described by Rosenblatt, 1958. This “perceptron algorithm” was a binary classifier that connected the input neurons (one for each input variable = feature) to an output neuron (response). If the signal in the output crossed a certain threshold (activation function), the predicted class changed (e.g. from ‘0’ to ‘1’). However, because of its limited flexibility, and particularly its inability to represent nonlinear relationships, the perceptron fell into oblivion for many years until it was discovered that additional layers between the input and outout neurons (so-called ‘hidden’ layers) made it possible to fit any type of function shape (see subsection “deep learning”). The added flexibility is achieved by the hidden layers in conjunction with the non-linear activation function: as in the brain, a hidden neuron will only ‘fire’ if the accumulated signals of the previous layers surpass a certain threshold. This structure allows ANNs to flexibly approximate practically any functional shape. The potential of ANNs for ecological applications was recognized early (Foody, 1995; French & Recknagel, 1970; Simpson et al., 1992), although to today they have largely been replaced by the more advanced Deep Neural Networks in E&E .

## Deep learning

DL models represent the latest methodological advance in ML (Fig. 2). DL algorithms are neural networks, with the difference that they include a large number of hidden layers (Borowiec et al., 2022; LeCun et al., 2015) and that the arrangement of the hidden layers (= architecture) is often different from the simple, fully connected ANN. Complex task-specific architectures, often with millions of parameters and specific structures, evolved over the years (for example residual neural networks (He et al., 2016), see also Table 1).

Although DL is based on the same ideas and principles as all other ML algorithms, it is commonly treated as a novel field because of its distinct principles (see section "Why ML works") and the task-specific architectures that do not resemble traditional ML models. For example, convolutional neural networks (CNNs) are a special architecture that is commonly used for image-based tasks (e.g. species identification, see Borowiec et al., 2022). Recurrent neural networks (RNNs) are another architecture that is applied for time series tasks (Table 1; Christin et al., 2019; LeCun et al., 2015). In ecology, RNNs have been used, for example, to predicting population dynamics (Joseph, 2020) or animal movements (Rew et al., 2019) (see Borowiec et al., 2022 for more details about different DL algorithms). DL algorithms have also been used to synthesize taxonomic information from literature (Le Guillarme & Thuiller, 2022), predicting species interactions in ecological networks (Strydom et al., 2021), predicting species distributions (Deneu et al., 2021), identifying species (Ferreira et al., 2020), automatic monitoring of species (Norouzzadeh et al., 2018; Tuia et al., 2022), or landscape classification (Stupariu et al., 2022). In the following, we treat DL as a subfield of ML and will mention DL only when relevant differences to classical ML algorithms are involved.

## Why does ML work?

When considering current DL algorithms with millions of parameters, researchers trained in classical statistics often struggle to understand why they should work at all. A statistical model with a similar number of parameters could likely not even be fit on a reasonable amount of data (e.g. in a linear regression model, if p (number of parameters) is larger than n (number of observations), there are no more degrees of freedom and the equation system is underdetermined). And even if fitting the model was possible, the bias-variance trade-off that is fundamental to both statistics and ML (Box 3, Fig. IIIa) predicts that the optimal compromise between systematic model error (bias) and error due to variance (parameter uncertainty) is at intermediate model complexity, which means that a simpler, alas biased model often has a lower total error than a complex, more unbiased model (Box 3, 4). Excessively large models should therefore overfit the data and generalize badly.

## Box 3: Generalization error and the bias-variance tradeoff

By making models more complex, one can make the predictive error on the training data (**in-sample error**) arbitrarily small. What we really care about, however, is the ability of a model to predict to new (out-of-sample data). The discrepancy between model predictions and observations on independent data ( e.g. generated by an appropriate cross-validation, see Roberts et al., 2017) is called the **generalization error**.

When minimizing the generalization error, there exists a fundamental tradeoff between variance (model uncertainty) and bias. More complex have higher variance, but also lower bias because (Fig. IIIa). The two counteracting errors usually lead to a sweet spot of the generalization error at intermediate model complexity. Interestingly, DL models display double sloping curve of generalization error (Fig. IIIb), suggesting that the variance of deep neural networks does not increase for very wide and deep networks. The reasons for this are still discussed in the literature.

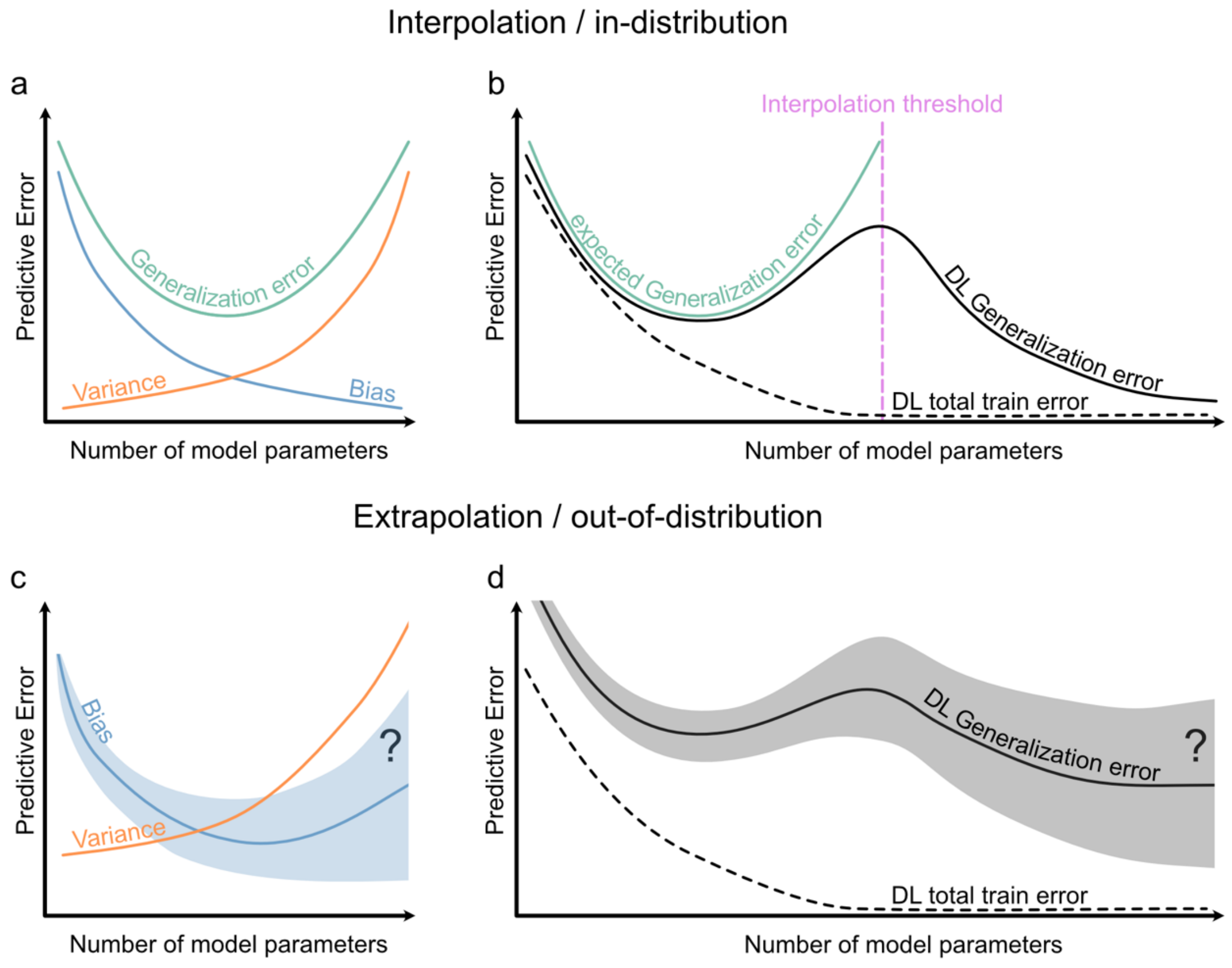

**Figure III:** Typical bias-variance trade-offs in classical machine learning (left panels) and deep learning (right panels) models for interpolation (in-distribution) and extrapolation (out-of-distribution) tasks. Note in panel b) that after the interpolation threshold (pink dotted line) the training loss is constant (i.e. bias is not improved

by increasing model complexity), but the test loss (and thus variance) can be still lowered by increasing the model size. Also, note that the total generalization error in extrapolation tasks (panels c and d) is usually higher and the optimal model complexity lower, as the bias will not go to zero with increasing model complexity (depending on the similarity between training and test data).

**Overfitting / Underfitting** describes a situation where the generalization error is higher than necessary or expected. When considering interpolation (in-distribution, Fig. IIIa, b) tasks, overfitting/underfitting is usually linked to too high/low model complexity, which leads to a poor compromise between bias and variance. In extrapolation (out-of-sample, see Fig. IIIc, d) tasks, the reasons for overfitting are often more rooted in bias problems, meaning that the patterns learned in the training data do not generalize to the test data (see, e.g. Yang et al., 2020 as an example for vision tasks).

Despite that, the practical experience is that ML models converge and generalize well to new data (at least to in-distribution predictions; out-of-distribution predictions are as challenging for ML as for other approaches), suggesting that they do not overfit. Not only that, if complexity is measured by the number of parameters, it was observed that the bias-variance trade-off for deep neural networks actually follows a double-descent curve (Box 3, Fig. IIIb)(Belkin et al., 2019; Nakkiran et al., 2019): beyond a certain point, increasing model parameters, different to our expectation, decreases generalization error, indicating that overparameterization can even be beneficial for reducing total predictive error (Nakkiran et al., 2019). Indeed, making neural networks both deeper and wider can reduce the generalization error (Arora, Du, et al., 2019; Huh et al., 2021; Novak et al., 2019; Shwartz-Ziv & Alemi, 2020), indicating that the number of parameters in DL is a poor measure of effective model complexity (Birdal et al., 2021), which is confirmed by more appropriate complexity measurements (Box 3, Fig. IIIb) (Birdal et al., 2021; Nakkiran et al., 2019).

The answer to this conundrum is that practically all ML approaches, despite having formally a very high number of parameters and the associated ability to model complex input-output relationships, perform implicit complexity adjustments that limits their flexibility and avoids overfitting. We divide these optimizations into two categories: internal (algorithmic) and external (optimization) based approaches.

## Internal (algorithmic) complexity optimization

By internal (algorithmic) complexity optimization, we understand algorithmic structures that lead to a self-adaptation of model complexity.

One basic mechanism to generate this behavior is that many ML algorithms implicitly or explicitly generate predictions from an ensemble. A model ensemble may include many parameters, but the effective complexity of such an ensemble is by no means the sum of the complexity of the ensemble members. Rather, it is typically more in the order of the average complexity in the ensemble, but also depends on the difference between the ensemble members, which in turn affects error and variance of the ensemble estimator (Bernardo & Smith, 2009, Dietterich, 2000; Dormann et al., 2018; Ganaie et al., 2021). Because ensemble members are fit to data, the data can influence the difference between ensemble members and thus the complexity of the total ensemble estimator.

To support this behavior, many ML algorithms include (tunable) mechanisms to increase variance in the ensemble. For example, bagging decreases similarity between ensemble members by using bootstrapping the data (Sagi & Rokach, 2018). RF goes one step further by using a random subset of the features in each node, which further diversifies and decorrelates the individual models (Breiman, 2001a) and decreases the variance of the ensemble (Breiman, 2001). In gradient boosting (see Friedman, 2001), the subsequent trained models depend on the previous model but they are uncorrelated because the following members are forced to compensate the errors of the previous model (Sagi & Rokach, 2018).

## External (optimization) adaptation of model complexity

On top of internal mechanisms to adopt model complexity, most practical ML pipelines apply an additional optimization step where hyperparameters of the model are optimized under cross-validation (or simply into train, development, and test splits for large DL models).

Hyper-parameters are parameters that do not directly control predictions, but rather the architecture (e.g., number of nodes of a of neural networks or the number of trees in a RF) or the learning behavior of ML algorithms. Some ML algorithms have few (e.g., RF), others many (BRT or DL) hyperparameters. Hyper-parameters are usually tuned via a nested cross-validation setup, i.e. an outer cross-validation to estimate generally the predictive errors of the model and an inner cross-validation which controls the tuning (see Table 2 for ML frameworks).

A particularly important class of hyper-parameters are regularization parameters, which control the flexibility of the algorithms. In general, regularization means that constraints are imposed on an algorithm to limit its flexibility. The type and strength of regularization depends on the task, data, and the algorithm, but the most common regularization type is a so-called shrinkage penalty which bias parameter estimates to a certain value, typically zero. For example, L1 (LASSO (Tibshirani, 1996)) and L2 (Ridge) (Hoerl & Kennard, 1970), or elastic-net if combined (Zou & Hastie, 2005) biases the estimates intentionally to zero. Shrinkage penalties were originally developed to estimate complex statistical models (e.g., if number of observations << number of predictors) such as linear or logistic regression models but have been adopted in ML models ever since. In tree-based methods (e.g., RF), hyper-parameters such as the depth of the trees have regularizing effects, whereas in DL a range of regularization techniques is used, for example L1 or L2 on weights and dropout (where random parts of the network are set to zero during training, see Srivastava et al., 2014).

## Open questions regarding model complexity in deep learning

While the principles of internal and external complexity adoption are central to both classical ML and DL algorithms, it is usually assumed that, by themselves, they are not sufficient to explain the success of the highly complex DL algorithms. A particularly puzzling observation is that in DL algorithms, generalization improves with model, even after the training loss has reached a value close to zero (Fig 2b), a behavior that is not observed in simpler ML algorithms.

One hypothesis to explain the discrepancy is that overparameterization in combination with stochastic training of the networks (stochastic gradient descent) leads to an implicit regularization (Arora, Cohen, et al., 2019; Huh et al., 2021; Li et al., 2021). This would explain why deep neural networks display a bias towards simpler functions (De Palma et al., 2019;

Valle-Pérez et al., 2019) that increases with the depth of the networks (Huh et al., 2021), independent of using deeper or wider networks (Nguyen et al., 2021). Moreover, it was observed that often over 90% of the trained networks' parameters can be set to zero with little or no loss of generalization accuracy (Frankle & Carbin, 2019). Such a pruning can reduce the computational cost of the model and lower the generalization error (Bartoldson et al., 2020) or identify robust models (Kuhn et al., 2021). It was also suggested that the random initialization of a large DNNs is more likely to create a good subnetwork (Frankle & Carbin, 2019; S. Zhang et al., 2021), which is then identified by training, regularization or pruning (S. Zhang et al., 2021).

This does not fully answer the question about the superiority of DL but it provides at least some clues. Future work will show how or what exactly DL learns. Moreover, most modern DL archichtectures consist not of pure fully-connected layers anymore but are a mixture of different layer types and can even use common ML concepts such as boosting or bagging. For example dropout generates theoretically an infinite number of subnetworks and is similar to an ensemble model (Srivastava et al., 2014). Deep residual networks (He et al., 2016) for image-based tasks resemble boosting and thus ensemble models (Veit et al., 2016).

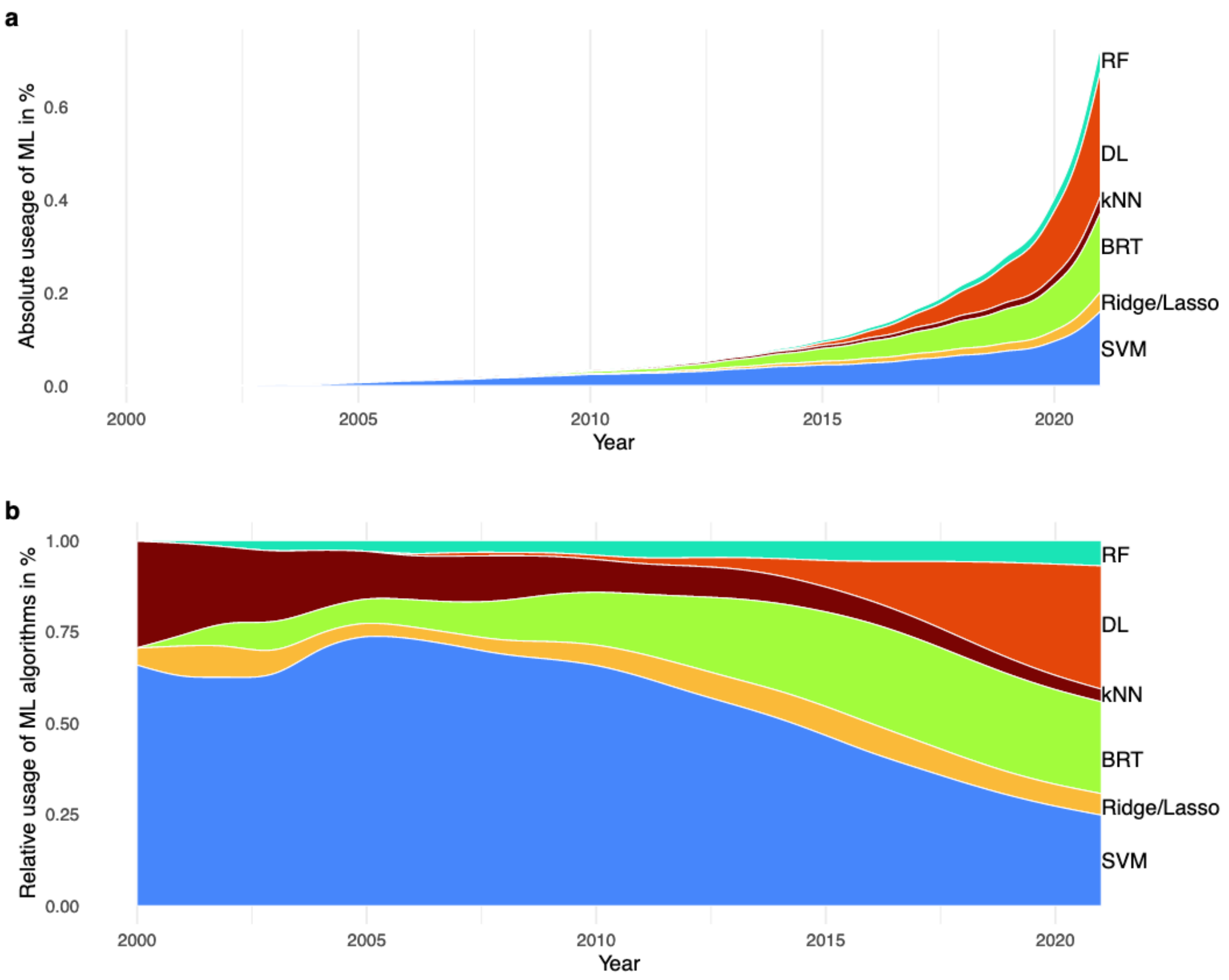

**Figure 3:** Development usage of ML algorithms (RF, DL, kNN, BRT, Ridge/Lasso, SVM) in literature from the E&E field (see Appendix S1.1 for more details about the trend analysis). Panel a) shows the absolute change in their usage in percent and panel b) shows the relative change in their usage in percent. The overall usage of ML models increased strongly over the last twenty years and especially DL attracted a lot of attention in the last ten years. (SVM = support vector machines; RF = Random Forest; BRT = Boosted regression trees; Ridge/Lasso = ridge or lasso or elastic-net (ridge and lasso) regression; DL = Deep Learning; kNN = k-nearest neighbor).

## Emerging Trends in ML (in E&E)

In the last section of this paper, we want to look at current trends in ML, and speculate how they will impact the E&E field.

### Trends in algorithm use in E&E

To understand what ML and DL algorithms are used by ecologists, we performed a text analysis of the E&E literature over the last decades (details see Appendix S1.1). Our results show that the use of both ML and DL methods in E&E increased sharply in the last decade, but classical ML methods still dominate in practical applications (Fig. 3). Of those, SVMs were the most popular algorithm in the early 2000s, but lost dominance since. BRTs became popular during the mid 2010s (Fig. 3B), and more recently, neural networks (including DL) are rising in popularity (Fig. 3B). The increase in publications using DL approaches explains why so much attention is being paid to these algorithms in recent reviews, but our analysis (Fig. 3B) also highlights that proportionally, classical ML methods remain important in the future.

We anticipate that classical ML will probably remain important future, as many tasks in E&E are more naturally approached with simpler ML algorithms. In particular, there is little evidence that DL can outperform classical ML algorithms on supervised learning on a limited amount of structured (tabular) data (cf. Strydom et al., 2021), simply because their higher flexibility, which also necessitates hyperparameter tuning, only creates an advantage if the data is large and complex enough. Thus we believe that classical ML algorithms will remain important, for example, for species distribution modelling (Beery et al., 2021; Elith & Leathwick, 2009), for example for identifying conservable or restorable areas (Moradi et al., 2019; Cheng et al., 2018; Duhart et al., 2019; Kwok, 2019); forest management (Lauer et al., 2020); ecosystem service management (Dietterich et al., 2012; Scowen et al., 2021); wildlife management (G. R. Humphries et al., 2018) and conservation (see Tuia et al., 2022); assessing the risk of invasive species (Barbet-Massin et al., 2018; Jensen et al., 2020); and biodiversity assessments (Distler et al., 2015). Other tasks where classical ML will likely remain competitive include filling (knowledge) gaps in datasets (Penone et al., 2014) or in ecological networks (e.g. food webs (Desjardins-Proulx et al., 2017), plant-pollinator networks (Pichler et al., 2020a), host-parasite networks (Dallas et al., 2017, 2021)) and predicting potential wildlife hosts of zoonotic diseases (Albery et al., 2021; see Becker et al., 2022; Han et al., 2015; Wardeh et al., 2021).

DL algorithms, on the other hand, will likely further rise in popularity for analyzing complicated and unstructured data in E&E, for example for the identification of species in aerial images (Ferreira et al., 2020; Gray et al., 2019; Guirado et al., 2018; Torney et al., 2019) or camera (trap) images (Ferreira et al., 2020; Mäder et al., 2021; Tabak et al., 2019; Willi et al., 2019; Fritzler et al., 2017; Lasseck, 2018; Stowell et al., 2018; Beery et al., 2020; Norouzzadeh et al., 2021; Van Horn et al., 2018; Mac Aodha et al., 2018; Fairbrass et al., 2018).

Clustering and ordination methods have a long tradition in ecology and ML algorithms dominate many unsupervised learning tasks (Box 2) such as species delimitation (Derkarabetian et al., 2019), outlier detection, identification of eco-provinces (Sonnewald et al., 2020) or operational taxonomic units (OTUs) in metabarcoding (Deiner et al., 2017). Here, however, DL can expand clustering or dimension reduction to unstructured data such as images in remote sensing (Zerrouki et al., 2021) or (genomic) sequences (D. Wang & Gu, 2018).

As more and more data will become available in the future (Albery et al., 2021) and ecologists have just started to grasp the applications of deep learning, we anticipate that these applications will continue to grow faster than classical ML algorithms, until the time that a methodological equilibrium is reached.

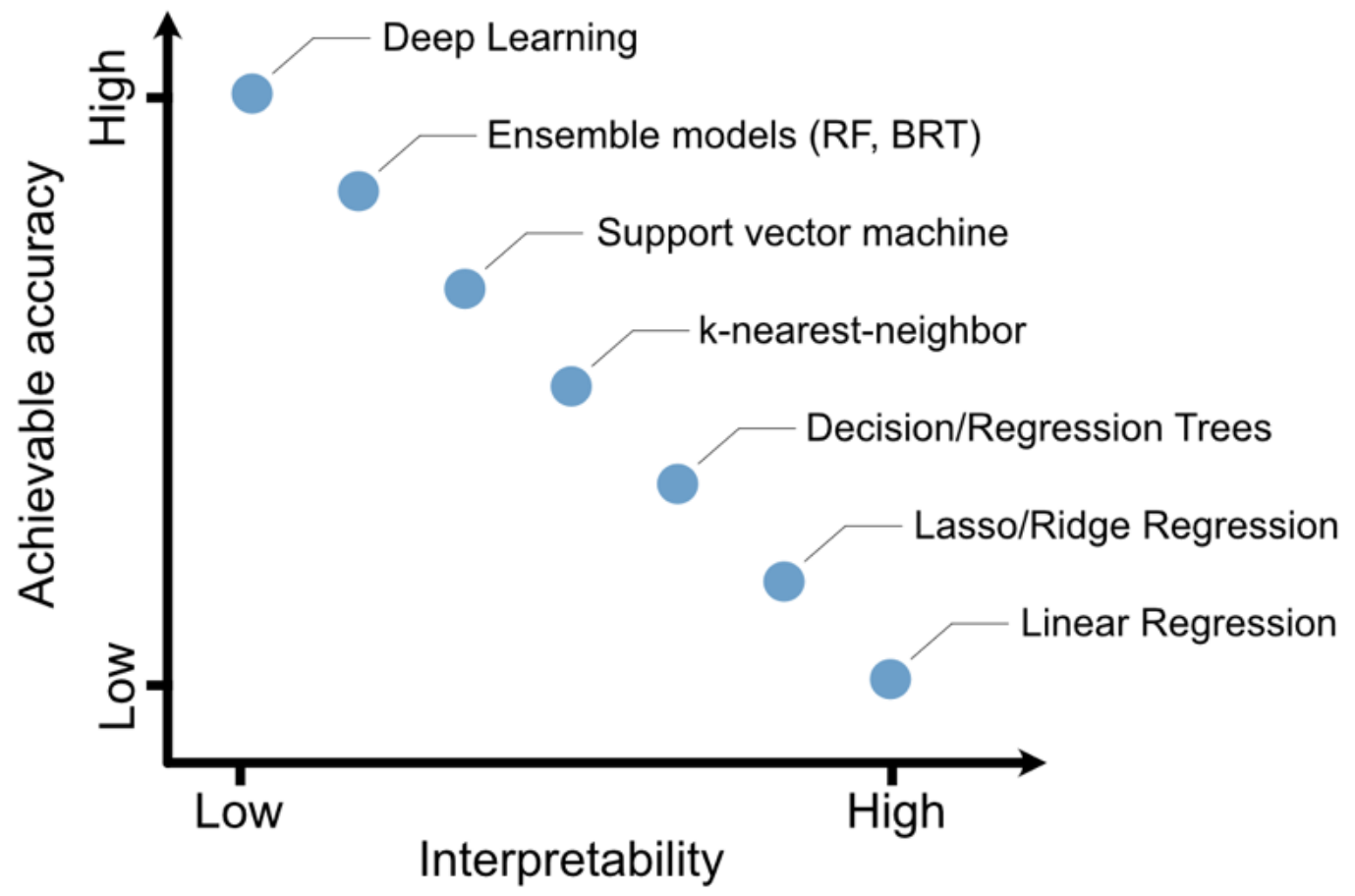

**Figure 4:** Conceptual illustration of the trade-off between achievable accuracy and interpretability of ML and DL algorithms. DL can achieve the highest accuracy but shows the lowest interpretability.

## New applications for ML in E&E

Moreover, when anticipating the future of ML and DL algorithms in E&E, we expect that their use will be expanded beyond classical prediction and classification tasks on ecological data. For example, Davies et al. (2021) demonstrated that DL can aid researcher by generating new hypotheses which were tested afterwards. Another interesting field for ML is simulations and simulation-based inference. For stochastic simulations or big process-based models, likelihoods are often intractable or computationally expensive to evaluate (e.g. phylogenetic analyses). ML and DL algorithms can support simulated-based inference by generating new summary statistics (e.g. Hauenstein et al., 2019), by being incorporated in process-based models for computational gains (e.g. Rammer & Seidl, 2019), or by emulating them (S. Wang et al., 2019). Moreover, ML can also be used to predict the parameters of complex stochastic models (Voznica et al., 2021, Roy et al., 2022), and thus act as a likelihood-free calibration method, similar to approximate Bayesian computation (Hartig et al., 2011).

Moreover, in the era of cheap sensors and other data collection sources, the dimensionality of the data is often difficult to handle with traditional methods. Unsupervised learning algorithms can help to reduce data dimensionality and detect pattern and trend, for example before data is used in downstream supervised learning tasks (Strydom et al., 2021; Zerrouki et al., 2021), to handle the data itself (Alves de Oliveira et al., 2021), or to detect anomalies in the data (J. Zhang et al., 2021).

## Rethinking the data collection process in the light of the new methods

The wide availability of DL algorithms could also have a strong impact on data collection in E&E. Image recognition methods can reduce labor costs and thus help to create much larger datasets. DL can identify species in different data types (Ferreira et al., 2020; Gray et al., 2019; Guirado et al., 2018; Torney et al., 2019; Mäder et al., 2021; Tabak et al., 2019; Willi et al., 2019) or extract information such as traits from raw data (Dunker et al., 2020). Moreover, technical advances in eDNA and other sensor data allow to produce much larger datasets (e.g.

see Pimm *et al.* 2015) which can then be processed and combined by ML and DL algorithms (see Tuia et al., 2022).

One of the advantages of establishing such machine-assisted data collection and processing pipelines is that they are reusable by many researchers, similar to the development in sequencing technology. For example, once an image-based species recognition pipeline is established, it can be re-used without using the time from taxonomic experts for data analysis. So far, there are few examples of such ready-to-use pipelines for realistic data collection tasks, and those that exists do not always perform and generalize well to new situations. However, we think the field should develop ML models for data collection that are available to everyone (McIntire et al., 2022) and don't have to be retrained by experts. The NLP community demonstrate how it could be done: Model hubs with many different pre-trained models and a simple and common interface usable by everyone (e.g. Wolf et al., 2019; Mäder et al. 2021, cf. 2021; Ott & Lautenschlager).

## Making ML work with small data sets

A pervasive problem for the applications of many modern ML algorithms for E&E is that their training requires data sizes that are rarely available. New DL techniques such as few-shot learning (see Y. Wang et al., 2020) or transfer learning can reduce the necessary amount of data enormously, and are thus particularly interesting to ecologists. As an example, most DL-based image classifiers consist of two stages: first, they identify edges and shapes in the images, and secondly, they classify the shapes (LeCun et al., 2015). The first stage makes up a major part of the model and is both data and resource-intensive. Research has shown that this part of the network is relatively generic, and usually only the second stage has to be re-trained when adopting a network for a new task (Weiss et al., 2016; Zhuang et al., 2021). Thus, many big model architectures can be downloaded pre-trained and can then be fine-tuned for a new task (transfer learning).

Options such as transfer learning, however, apply mainly for vision-based tasks and DL, and not to classical structured tabular data. In the latter, the response is often explained by specific relationships with a particular feature, which rarely generalizes to other tasks. In such a situation, there is little to gain by applying transfer learning, which may also in part explain why DL rarely outperforms traditional ML algorithms on small classical structured tabular data (Pichler et al., 2020; Schwartz et al., 2020 ; cf. Arik & Pfister, 2020).

Small datasets are common in E&E because often observations are difficult to sample (e.g. ecological networks (Maglianesi et al., 2014; Strydom et al., 2021) and because ecological patterns can differ over scales (Poisot et al., 2015) datasets can rarely be combined. However, this is where E&E can benefit from the wide range of different ML algorithms (Fig. 4): SVMs or kNN can handle sparse datasets well (Como et al., 2017; Drake et al., 2006) and LASSO, Ridge, and elastic-net regressions are well suited for datasets with more features than observations (Zou & Hastie, 2017). Also, the functional form of the correlation between features and the response determines the complexity required. If simple, less complex models can already achieve a high accuracy (e.g. Pichler et al., 2020) but have a better tradeoff in terms of interpretability and usability. However, the relationship is difficult to infer a priori which explains the need to compare different algorithms for the same task (Faisal et al., 2010; Norberg et al., 2019; Pichler et al., 2020).

## Transparency and bias of decision based on ML and DL models

Predictions and research in E&E often intersect with policy and decision-making (de Groot et al., 2010; Sofaer et al., 2019).  ML models are increasingly used in this context, for example for conservation planning (Huettmann, 2018), management decisions (G. R. Humphries et al., 2018), agriculture management (e.g. crop management)(Liakos et al., 2018), and disease control (e.g. Romero et al., 2021). We thus anticipate that problems of bias and transparency will emerge, as they have in other fields were ML was used to make decisions relating to human welfare (e.g. Hardt et al., 2016, Vayena et al., 2018).

Transparency refers to the issue that stakeholders may question why certain predictions are being made and thus if they can trust these predictions. Without satisfying answers to these questions, ML decisions may also be legally challenged. Although it is not impossible to answer these questions for ML models (see next subsection on xAI), it is doubtlessly more challenging to understand and communicate the logic of ML decisions, compared to simple statistical models (Fig. 4).

This lack of transparency also makes it difficult to understand if an algorithm exhibits bias. In this context, bias is understood more broadly than in statistics, and includes both the use of non-representative or socially undesirable training data and the use of features that should normally not be used in decisions (e.g. gender, race). The former occurs if the training data was disproportionately collected in different groups or regions (J. Zou & Schiebinger, 2018) and isn't representative of what the algorithm should learn. A common example from social science is that algorithms trained on classical literature or texts often learn gender-biased word associations, such as a preference for doctors to be male. Biased data is a significant problem for E&E since geographic (Martin et al., 2012; C. Meyer et al., 2016) and taxonomic (Pyšek et al., 2008; Trimble & van Aarde, 2012) sampling biases are common. The latter describes situations where the data may be representative, but certain features should not be used for ethical reasons. The challenge here is that these features may be implicitly encoded by other features and thus be used in ML and DL algorithms, even if they are not explicitly provided in the data (e.g. ethnic background can be inferred from the people's urban districts (Feuerriegel et al., 2020; Hardt et al., 2016)) (Caliskan et al., 2017).

To understand these issues and to find solutions, the field of responsible and trustworthy AI has formed at the intersection between AI and social science disciplines (sociology, psychology, law). The focus of responsible AI is on creating fair and sustainable ML and DL models and avoid their misuse or misinterpretation (e.g. Barredo Arrieta et al., 2020; Wearn et al., 2019). For example, it is possible to algorithmically detect biases (Cirillo et al., 2020) and correct the models accordingly (e.g. Alvi et al., 2018; Kim et al., 2019) by using fairness metrics to guide training so that underrepresented groups are not neglected (e.g. Liu & Vicente, 2021).

## Explainable AI as peephole in black-box models

The previously mentioned transparency issues arise because ML algorithms become increasingly difficult to interpret with rising model complexity (Fig. 4, Breiman, 2001b).

Although some algorithms provide metrics for feature importance (e.g. Breiman, 2001a; Friedman, 2001), ML algorithms usually do not provide simple effect estimates, nor do they provide measures of certainty, such as confidence intervals or p-values. This not only poses a problem for ethical transparency, but also for researchers that would like to understand why predictions are made for scientific reasons.

To solve this problem, the field of explainable AI (xAI) has emerged that develops tools to understand how ML models make their predictions. Most xAI methods are post-hoc, meaning that the model is first trained and afterwards investigated (Barredo Arrieta et al., 2020; Lucas, 2020). xAI differs from other similar-sounding approaches such as identifying predictive trait profiles (domain expertise is used to group features and by including and removing them from the model, their predictive capabilities are estimated (e.g. Han et al., 2015)), in that the goal is to understand the model itself because of potential downstream applications (e.g. by using the predictions). Global xAI methods try to summarize the models by generating variable importance (similar to the natural variable importances from RF and BRT ) (Fisher et al., 2018) or simplify the models by approximating the original model with interpretable models (Molnar, 2020). Local xAI methods try to explain single predictions (Ribeiro et al., 2016). In ecology and evolution, for example xAI methods are already used to assess trust of predictions from SDMs (Ryo et al., 2021). Although there are still many questions about the reliability of these methods, in particular under collinearity of features (Hooker & Mentch, 2019; Yu et al., 2021, but see Apley, 2016), xAI is becoming an indispensable tool for working with ML models, and presumably there will be specialized xAI methods for ecological applications.

## Causal inference with ML

Finally, it is a common place that models can give us the right answer for the wrong reason, and sometimes it is even easier to get good predictions for the wrong reasons. For example, when several features are collinear, including all of them increases the uncertainties of the estimates (Greenland, 2003; Lederer et al., 2018), which negatively impacts the predictive performance (Dormann et al., 2013; Hoerl & Kennard, 1970). Thus, excluding collinear features may improve predictions, even though this destroys the causal structure and means that the models predict the right things for the wrong reasons (Arif & MacNeil, 2022). Interestingly, some of the regularization techniques now widely used in ML were originally developed to reduce collinearity problems (Lederer et al., 2018), although that doesn't mean that they neccesarily always 'select' the causal one from two collinear features (Zou & Hastie, 2005).

## Box 4: Predictive versus causal model building strategies

The best predictive model doesn't need to correspond to the true causal model. To demonstrate this, we created a simulated dataset, where the response variable Y is affected by the feature X with a causal effect of 0.5 and by a second feature C with the causal effect of -1.5 (Fig. IV). C also has a causal effect of 1.0 on X. X and C are thus highly correlated (> 0.9 Pearson correlation factor). We fitted different models (full model, model with only X or C, and full model with a ridge regularization (lambda = 0.01), Fig. III) on 20 observations and estimated the predictive error (root-mean-squared-error, (RMSE)) on 480 observations of the holdout. We repeated the simulation and the model fitting 10,000 times.

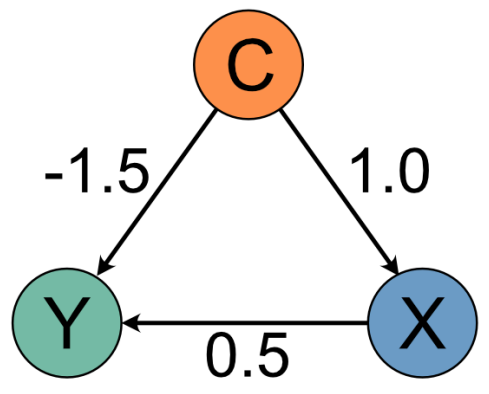

**Figure IV:** Small example. We are interested how X effects Y. C is a confounder, i.e., affecting X and Y. C and X are highly correlated (> 0.9 Pearson correlation factor). Numbers show the true effect estimates

| Model | X-estimate | C-estimate | RMSE on holdout |
|---|---|---|---|
| Y ~ X + C | 0.5 | -1.5 | 0.109 |
| Y ~ X | -1.0 | . | 0.107 |
| Y ~ C | . | -1.0 | 0.106 |
| Ridge | -0.32 | -0.59 | 0.106 |

**Figure V:** Results of different model specifications. We simulated 10,000 times from our small example (Fig. I) and fitted four different models on 20 observations: a) The full model with the Confounder (C) and our variable of interest (X), b) only our variable of interest, c) only the confounder, and c) the full model with a ridge regularization (lambda = 0.01). We evaluated the predictive error of the models by calculating the root-mean-squared error of the predictions for the holdout (480 observations).

**In causal inference**, the objective is to obtain correct effect estimates, which means that sometimes otherwise uninteresting collinear features have to be included to control the confounding (variable C, Fig. IV), while other structures, such as collider, most be excluded to obtain correct estimates. Thus, in causal analysis, the focus is to establish a correct hypothesis about the causal structure to obtain correct effect estimates (first model, Fig. V). The causal structure can be based on logical considerations or causal discovery algorithms. Importantly, minimizing predictive error is not the primary goal of the analysis, and controlling confounders often increases uncertainties of the parameter estimates that propagate through the model and negatively affect the predictive error (note that the true causal model shows highest RMSE, Fig. V).

**In predictive modeling,** our goal is to minimize the predictive error of our model. A common strategy is to provide the model with all the variables and use methods such as regularization or AIC selection to reduce model complexity and find the sweet spot of the bias variance trade-off (see Fig. V last model). In such an approach, collinear features will often be removed because they increase uncertainties while contributing relatively little to the prediction, given that their effects can be “emulated” by other features. In our simulation (Fig. IV), we see that the true causal model has the highest predictive error, but correct effect estimates. The other models, have incorrect estimates but smaller predictive errors.

The explicit selection of ML for predictive performance means that we should be skeptical about whether predictions are made based on the correct causal structure (Box 4). This also means that we should be careful to interpret xAI methods as “effect estimates” – if a certain feature is strongly used by the ML model, it could be because this feature has a biological or ecological effect on the response, but it could also easily be because it correlates with other features (e.g. Genuer et al., 2010). Statistical control (e.g. for confounders) with ML models is far more difficult than with simple linear regressions.

Nevertheless, there are interesting ideas to exploit ML for (causal) data analysis. For example, if we have a high dimensional dataset which would be difficult to analyze with conventional statistical tools, we could use ML and xAI to identify interesting patterns or features (Lucas, 2020; Pichler et al., 2020), and test those later in a confirmatory analysis.

Second, based on a causal analysis, ML may be used to achieve more exact control for confounders (e.g. Joseph, 2020; Tank et al., 2021; Wein et al., 2021; Zečević et al., 2021) or to combine statistical models with ML and DL algorithms (Masahiro & Rillig, 2017; Tank et al., 2021). A third idea is to incorporate at least causal constraints into ML are physics induced (or informed) neural networks (PINN), where physical laws are integrated as constrains into the learning of neural networks (Karniadakis et al., 2021, 2021; Rackauckas, Edelman, et al., 2021). PINNs were originally developed to improve predictions but could also mark an important milestone in combining ML and causal inference by forcing the models to adhere to known causal constraints rather than using a completely naïve, data-driven model that ignore functional knowledge about the systems. This might be an interesting approach for a field as E&E that has acquired a lot of knowledge about the different systems over the years. Finally, there is active research to create ML methods that directly achieve causal discovery. For example, DL research has shown promising advances in symbolic regression where equations for systems are automatically inferred (Cardoso et al., 2020; d'Ascoli et al., 2022).

In summary, causal inference or causal discovery with ML is challenging, but there are various ways to combine machine learning methods with causal inference and based on the interest in this topic in the ML and DL field, but also the interest of ecologists to understand causal relationships in their data, we believe that the importance of this topic will increase in the coming years.

## Conclusion

ML and DL algorithms are powerful tools for predictive modeling and data analysis. While DL algorithms have conquered image-based tasks, classical ML algorithms such as RF and BRT still excel on structured data. The superior performance of ML and DL algorithms compared to statistical models can be explained by their higher flexibility and automatic data-dependent complexity optimization. For example, ML algorithms such as RF and BRT balance complexity by combining weak uncorrelated models into an ensemble, and DL uses a combination of indirect regularization through overparameterization and stochastic gradient descent. Compared to classical statistical tools in E&E, ML methods are rather optimized for prediction, and caution is advised with their causal interpretations (Box 4). Nevertheless, ML and DL have become an indispensable tool in E&E for data processing, predictive modeling, and data analysis, and we anticipate its use in E&E studies will only increase in the future.

When forecasting, a challenge shared by classical statistical models and ML alike is making predictions that extrapolate outside the feature space used to train the model (out-of-distribution predictions, see Beery et al., 2018; Koh et al., 2021; Box 3). The reason why such predictions often fail is that predictive models often learn to use non-causal proxies, but also that relationships do not necessarily remain linear outside the area of data. Efforts have been made to identify such potentially unreliable predictions a priori (e.g. Meyer & Pebesma, 2021) or to correct for them (e.g. Tseng et al., 2022). Possibly, also the inclusion of ecological or mechanistic understanding, for example as additional model constraints, could help to improve model generalizability. However, the problem of extrapolation is fundamental to any data-driven prediction approach, which should be kept in mind when using ML for such predictions (Box 3).

Much research of the last years has focused on making ML algorithms more transparent and bridging the gap between the properties of classical statistical tools and ML tools (e.g. xAI). New methods such as Bayesian neural networks paved the way to obtain uncertainty and prediction intervals for DL models, closing the gap to statistical models (Ashukha et al., 2021; Loquercio et al., 2020). On the other hand, their use for causal inference is still disputed. While xAI allow to perform inference, the problem of causal inference with ML and DL is more fundamental: It is still unclear, for example, when and if they can distinguish between causal and collinear patterns which is a basic requirement for causal inference.

Despite the deserved attention, ML offers no free lunch. The focus of ML methods on predictions creates challenges for the interpretation of these models, and researchers will often be tempted to misinterpret them as causal (Box 4). Even more strongly than statistical models, ML depends on the quality and the quantity of the data. Because of this, we should carefully consider if the application of ML or even DL is necessary or promising for a task when simpler models with the advantages of higher interpretability, higher statistical power, and lower computational costs (and thus a better $CO_2$ footprint (Schwartz et al., 2020)) could do the job (Mignan & Broccardo, 2019). Nevertheless, we expect ML to become an indispensable tool in E&E, comparable to other traditional statistical tools such as linear regression models or analysis of variance models that have been used for many years.

## Acknowledgement

MP received funding from the Bavarian Ministry of Science and the Arts in the context of Bavarian Climate Research Network (bayklif). We would like to thank Daniel Rettelbach and Tankred Ott for their valuable comments and suggestions.

## Authors' Contributions

MP and FH jointly conceived and designed the study. Both authors contributed equally to the writing and preparation of the manuscript.

## Conflict of interest statement

The authors declare that they have no conflicts of interest.

## Data Availability

The analysis (trend analysis) and example code chunks for the different ML algorithms can be found for different programming languages (R, Python, and Julia) in the Supporting Information S1 or at https://github.com/MaximilianPi/Pichler-and-Hartig-2022.

# Supporting Information

## Trend analysis

For the global trend analysis in Figure 1, we used the R package ‘europepmc’ (v0.4.1, Jahn (2021)) to search from 1920 to 2021 the PubMed and Medline NLM databases. We used the following queries ‘deep learning’, (‘machine learning’ OR ‘machine-learning’), and (‘p value’ OR ‘p-value’ OR ‘statistically significant) as representatives for Deep Learning, Machine Learning, and classical statistical approaches. The number of hits were normalized by total hits in each year. For the stream charts in Figure 2, we used the search queries Table S1 and added them to the query’ AND (“ecology” OR “ecolog*” OR “evolution”) to restrict the queries to hits from the ecology and evolution field.

| Queries | ML and DL algorithm |
|---|---|
| (“artificial neural network” OR “deep neural network” OR “multi-layer perceptron” OR “fully connected neural network”) | Deep neural network (ANN) |
| (“convolutional neural network” OR “object detection”) | Convolutional neural network (CNN) |
| (“recurrent neural network”) | Recurrent neural network (RNN) |
| (“graph neural network” OR “graph convolutional”) | Graph neural network (GNN) |
| (“random forest”) | Random Forest (RF) |
| (“boosted regression tree” OR “boosted reg” OR “gradient boosting” OR “adaboost”) | Boosted Regression Trees (BRT) |
| (“k-nearest-neighbor”) | k-nearest neighbor (kNN) |
| (“ridge regression” OR “lasso regression” OR “elastic-net” OR “elastic net”) | Ridge, lasso, or elastic-net regression |
| (“support vector machine” OR “support vector”) | Support vector machine (SVM) |

Search queries and their corresponding ML and DL algorithm.

For the word clouds in Table 1, we used again the 'europepmc' R package to search abstracts and titles within the ML and DL algorithm specific queries (Table 1) for the following ecological keywords: species distribution, species interaction, mortality, remote sensing, invasive, decision making, ecosystem, species identification, species detection, extinction, functional trait, ecological network, biodiversity, and camera trap.

We used the R packages 'tm' (Feinerer, Hornik, and Meyer (2008)), 'wordcloud' (Fellows (2018)), and 'wordcloud2' (Lang and Chien (2018)) to analyze and create the final word cloud plots.

## Algorithms

A quick guide with code chunks in R, Python, and Julia for all common ML algorithms (elastic-net, SVM, kNN, RF, BRT, DNN, CNN, RNN, and GNN) can be found online at https://maximilianpi.github.io/Pichler-and-Hartig-2022/